\documentclass[twocolumn,secnumarabic,amssymb,nofootinbib,nobibnotes,aps]{revtex4-1}

\usepackage{graphicx}
\usepackage{bm}

\newcommand{\half}{{\textstyle{1\over2}}}
\newcommand{\thalf}{{\textstyle{3\over2}}}
\renewcommand{\vec}[1]{\bm{#1}}
\newcommand{\svec}[1]{\mbox{{\scriptsize \boldmath$#1$}}}

\begin{document}

\title{Dynamical generation of resonances in the P33 partial wave}

\author{B.~Golli}\email{bojan.golli@ijs.si}
\affiliation{Faculty of Education,
              University of Ljubljana, 1000 Ljubljana, Slovenia}
\affiliation{J.~Stefan Institute,
              1000 Ljubljana, Slovenia}
\author{H. Osmanovi\'c}
\affiliation{Faculty of Natural Sciences and Mathematics, University of Tuzla
              75000 Tuzla, Bosnia and Hercegovina}
\author{S.~\v{S}irca}
\affiliation{Faculty of Mathematics and Physics,
              University of Ljubljana,
              1000 Ljubljana, Slovenia}
\affiliation{J.~Stefan Institute,
              1000 Ljubljana, Slovenia}

\date{\today}

\begin{abstract}
We investigate the formation of resonances in the P33 partial wave
with the emphasis on possible emergence of dynamically generated 
quasi-bound states as a consequence of a strong $p$-wave pion 
attractive interaction in this partial wave, as well as their 
possible interaction with the genuine quark excited states.
By using the Laurent-Pietarinen expansion we follow the evolution
of the $S$-matrix poles in the complex energy plane as a function
of the interaction strength.
Already without introducing a genuine quark resonant state, 
two physically interesting resonances emerge with pole masses 
around 1200~MeV and 1400~MeV, 
with the dominant $\pi N$ and  $\pi\Delta$ component, respectively.
The added genuine resonant state 
in the $(1s)^3$ quark configuration
mixes with the lower dynamically generated 
resonance forming the physical $\Delta(1232)$ resonance, and pushes 
the second dynamical resonance to around 1500~MeV, which allows it 
to be identified with the $\Delta(1600)$ resonance.
Adding a second resonant state with one quark promoted to the 
$2s$ orbit generates another pole whose evolution remains well 
separated from the lower two poles.  
We calculate the helicity amplitudes at the pole and suggest that 
their $Q^2$ dependence could be a decisive test to discriminate 
between different models of the $\Delta(1600)$ resonance.
\end{abstract}



\maketitle

\section{Introduction}
Since its discovery in 1952~\cite{Anderson52a,Anderson52b},
the $\Delta$(1232) resonance has played a significant role in 
almost all attempts to understand the structure and dynamics of 
the nucleon and its excited states.
The early approaches were based on the observation that the 
dominant $p$-wave pion nucleon interaction is attractive in the 
P33 partial wave  --- in contrast to the P11, P13 and P31 
waves --- and may therefore generate a resonance at the observed 
energy, provided that the coupling is sufficiently strong. 
Since the pion-nucleon coupling constant was known from the
behavior of the P11 wave near the threshold, the above condition 
required a sufficiently large cutoff of the order of 
1~GeV/$c$~\cite{ChewLow56}.
With the introduction of the quark model the four charge states 
of $\Delta$(1232) have been identified as the isospin quadruplet 
belonging to the lowest quark decuplet.
The excitation energy with respect to the nucleon is usually 
explained by the gluon or/and pion exchange interaction between 
quarks.
The relatively strong $p$-wave pion-nucleon interaction  
--- though not the main mechanism to generate the resonance --- 
manifests itself in sizable pion contributions to the photo- and 
electroproduction amplitudes. 

While the properties of the $\Delta$(1232) resonance are well
understood, this is not the case with the next higher resonance
in the P33 partial wave, the $\Delta$(1600).
In the quark model, this resonance is traditionally described 
as the radial excitation in which one quark is promoted to 
the $2s$ orbit --- the analogue to the Roper resonance $N(1440)$ 
in the P11 partial wave. 
The problem with such an interpretation is that in the
harmonic-oscillator model the $2s$ excitation is twice as large
as the $1p$ excitation while the observed $N(1440)$ resonance
appears below the negative parity resonances.
Furthermore, recent results of lattice QCD in the P11 partial 
wave show no clear signal for a three-quark Roper state below 
1.7~GeV~\cite{lang16,kiratidis17}.
To resolve the problem of level ordering, an alternative 
approach has been proposed in which coupled-channel meson-baryon 
dynamics alone was sufficient to engender the 
resonance~\cite{Krehl,Roenchen13}.
In our previous work~\cite{PRC2018} we have shown that, while 
the mass of the $N(1440)$ resonance is indeed determined by 
the dynamically generated state with the dominant 
$s$-wave $\sigma N$ component, a genuine three-quark $(1s)^22s$ 
component with the mass above ~1750~MeV is needed to explain 
the properties of the resonance.
The presence of a bare baryon structure at around 1750~MeV has
also been emphasized in the EBAC 
approach~\cite{Sato-Lee10a,Sato-Lee10b}.

Since the $\Delta(1600)$ resonance may be considered as a 
spin $\thalf$, isospin $\thalf$ partner of the $N(1440)$, 
it seems at first glance that a similar model could work also
in the P33 partial wave with the $\sigma N$ substituted
by the $s$-wave $\sigma\Delta$ component.\footnote{%
In the following we shall denote the $\Delta(1232)$ as $\Delta$;
the $\Delta(1600)$ will be eventually denoted as $\Delta^*$.}
As shown in~\cite{PRC2018} the $N(1440)$ mass at the pole lies 
slightly below the nominal $N\sigma$ threshold, rather 
independently of the model parameters, which would mean that 
the mass of the $\Delta(1600)$ would be at least $200$~MeV higher 
than the mass of the pole given in PDG~\cite{PDG}, ruling out 
such a model.
Also, a preliminary calculation in the P33 partial wave 
has shown that the $\sigma\Delta$ component represents 
a rather minor contribution to the scattering amplitudes  
below $W\approx 1800$~MeV.
We therefore consider here an alternative model, based on 
the observation that the $\pi N$ as well as the $\pi\Delta$ 
interaction are attractive in the P33 partial wave, in which 
the dynamically generated state consists of quasi-bound 
$\pi N$ and $\pi\Delta$ states.
Such a model is further stimulated by the study~\cite{Roenchen14} 
using a semi-phenomenological approach to extract photoproduction 
couplings at the pole of $N$ and $\Delta$ resonances up to 
$W\approx2.4$~GeV, which has confirmed the dynamical origin of 
the $\Delta(1600)$ resonance with a dominant $\pi\Delta$ 
configuration.

There have not been many attempts to study the properties of 
$\Delta(1600)$ from the quark modeling point of view; let us 
mention the calculations in the relativistic quark model in 
a light-front framework~\cite{Capstick95,Ramalho10,Aznauryan15}
assuming the dominant $(1s)^22s$ quark configuration which 
leads to a similar behavior of the helicity amplitude as 
in the Roper case, and a calculation \cite{luya19} using 
a relativistic diquark-quark model.

In the next section we briefly review the basic features of 
our coupled-channels approach and of the underlying quark model
that has been used in our treatment of $N(1440)$.
However, in the version reported here we do not include the
$\sigma$ meson which turns out to have only a very minor
role in the relevant energy region.
Furthermore, since the $\pi NN$ coupling constant is well
established, we keep the $\pi$-quark coupling constant fixed 
and vary the cutoff parameter in order to study the evolution 
of the resonance  poles in the complex energy plane by using the 
Laurent-Pietarinen (L+P) 
expansion~\cite{L+P2013,L+P2014,L+P2015,L+P2014a}.
In Sect.~\ref{sec:scattering} we solve the coupled-channels 
problem, first without including any genuine three-quark resonant 
state, then by including a three-quark resonant state corresponding 
to $\Delta(1232)$, and finally adding a three-quark resonant state 
in a $(1s)^22s$ configuration.
In Sect.~\ref{sec:electro} we discuss the prediction of our model 
for the  photo- and electro-production amplitudes which may 
eventually support our picture of the $\Delta(1600)$.

\section{\label{sec:model} The model}

In our approach the scattering state in channel $\alpha$ 
which includes a quasi-bound quark state $\Phi_R$ assumes 
the form
\begin{eqnarray}
   |\Psi_\alpha\rangle &=& \mathcal{N}_\alpha\biggl\{
        \left[a^\dagger_\alpha(k_\alpha)|\Phi_\alpha\rangle\right]
   +  c_{\alpha R}|\Phi_R\rangle
\biggr.\nonumber\\ && \biggl. 
  +  \sum_\beta
   \int 
   {dk\>\chi_{\alpha\beta}(k_\alpha,k)\over\omega_\beta(k)+E_\beta(k)-W}\,
   \left[a^\dagger_\beta(k)|\Phi_\beta\rangle\right]\biggr\},
\label{PsiH}
\end{eqnarray}
where $\alpha$ ($\beta$) denotes either $\pi N$ or $\pi\Delta$ 
channels, [ ] stands for coupling to total spin $\thalf$ and 
isospin $\thalf$.
The first term represents the free pion and the baryon 
($N$ or $\Delta$) and defines the channel, the next term corresponds 
to a {\em bare\/} three-quark resonant state, while the third term 
describes the pion cloud around  the nucleon and $\Delta$.
Here $\mathcal{N}_\alpha=\sqrt{\omega_\alpha E_\alpha/(k_\alpha W)}$, 
$k_\alpha$ and $\omega_\alpha$ are on-shell pion momentum and energy, 
and  $W=\omega_\alpha+E_\alpha$ is the invariant mass.
The integral is assumed in the principal value sense.
The (half-on-shell) $K$ matrix is related to the scattering state 
as~\cite{EPJ2008}
\begin{equation}
   K_{\alpha\beta}(k_\alpha,k) = -\pi\mathcal{N}_\beta
  \langle \Psi_\alpha||V^\beta(k)||\Phi_\beta\rangle\,,
\label{eq4K}
\end{equation}   
with the property $K_{\alpha\beta}(k_\alpha,k) = K_{\beta\alpha}(k,k_\alpha)$.
It is proportional to the pion amplitude $\chi$ in~(\ref{PsiH}),
\begin{equation}
   K_{\alpha\beta}(k_\alpha,k)
       = \pi\,\mathcal{N}_\alpha\mathcal{N}_\beta\,
             \chi_{\alpha\beta}(k_\alpha,k) \,.
\label{chi2K}
\end{equation}

The amplitude $\chi$ satisfies a Lippmann-Schwinger type 
of equation:
\begin{eqnarray}
   \chi_{\alpha\gamma}(k,k_\gamma) 
   &=& -{c}_{\gamma R}\, V_{\alpha R}(k)
       + \mathcal{K}_{\alpha\gamma}(k,k_\gamma)
\nonumber\\ && 
+ \sum_\beta\int dk'\;
  {\mathcal{K}_{\alpha\beta}(k,k')\chi_{\beta\gamma}(k',k_\gamma)
  \over \omega_\beta(k') + E_{\beta}(k')-W}\,.
\label{eq4chi}
\end{eqnarray}
Our model utilizes two approximations for the kernel $\mathcal{K}$;
the first one implies only $u$-channel processes:
\begin{equation}
  \mathcal{K}_{\alpha\beta}(k,k') =  
  \sum_{i=N,\Delta} f^i_{\alpha\beta}\, 
  {V_{i\beta}^{\alpha}(k)\,V_{i\alpha}^{\beta}(k')
   \over \omega_\alpha(k)+\omega_\beta(k')+E_i(\bar{k})-W}\,,
\label{kernel}
\end{equation}
and the second one implies that the kernel can be made separable 
by assuming
\begin{eqnarray}
&&  \kern-30pt {1 \over \omega_\alpha(k)+\omega_\beta(k')+E_i-W} 
\approx    
\nonumber \\ &&
   {(\omega_\alpha + \omega_\beta + E_i - W)
\over 
   (\omega_\alpha(k)+E_i-E_\beta)(\omega_\beta(k')+E_i-E_\alpha)}\,,
\label{separable}
\end{eqnarray}
where $W = E_\alpha + \omega_\alpha = E_\beta + \omega_\beta$.
The factorization is exact if either of the $\omega$'s is
on-shell, i.e. $\omega_\alpha(k)\to\omega_\alpha=W-E_\alpha$ or 
$\omega_\beta(k')\to\omega_\beta=W-E_\beta$.
In the present work we include only pion loops and the nucleon 
and $\Delta$ as the $u$-channel exchange particles.
Based on our previous experience in the P11 and P33 partial waves 
these degrees of freedom dominate in the energy region considered
in the following.
The spin-isospin factors in~(\ref{kernel}) equal
$$
   f^N_{NN} = f^\Delta_{NN}={4\over9}\,, \quad 
   f^\Delta_{NN} = {1\over36}\,, \quad 
   f^\Delta_{\Delta\Delta} = {121\over225}\,,
$$
$$
    f^N_{N\Delta} = f^N_{\Delta N}={5\over9}\,, \quad
    f^\Delta_{N\Delta} = f^\Delta_{\Delta N}={2\over9}\,.
$$
Equation~(\ref{kernel}) implies dressed vertices; in the present 
calculation the vertices involving the $\Delta$ are increased 
by 30~\% with respect to their bare (quark model) values 
in accordance with our analysis of the P33 resonances 
in~\cite{EPJ2008}, while $V_{\pi NN}$ is kept at its bare value.

The vertices are determined in the underlying quark model
which can be chosen freely;
we use the Cloudy Bag Model~\cite{CBM} which involves two 
parameters, the pion-decay constant, $f_\pi$ (reduced to 76~MeV 
in order to reproduce the $\pi N$ coupling constant),
and the bag radius which determines the cutoff.
In our previous analysis we used a typical value of $R=0.83$~fm 
corresponding to the cutoff $\Lambda\approx 550$~MeV.
These two parameters describe consistently the scattering 
and photo-production amplitudes, including the production of 
$\eta$ and $K$ 
mesons~\cite{PRC2018,EPJ2008,EPJ2009,EPJ2011,EPJ2013,EPJ2016}.
In order to reveal the mechanism of $\Delta(1600)$ formation  
we study the evolution of the resonance properties as a function 
of $R$ (which is inversely proportional to the cutoff momentum) 
for a large range of its value, keeping in mind that the physically 
sensible interval  should be between 0.6~fm and 1~fm.

The pion amplitude can be written in terms of the resonant and
non-resonant part,
\begin{equation}
   \chi_{\alpha\gamma}(k,k_\gamma) = 
    c_{\gamma R}{\cal V}_{\alpha R}(k)+
    {\cal D}_{\alpha\gamma}(k,k_\gamma)\,,
\label{splitchi}
\end{equation}
such that~(\ref{eq4chi}) can be split into the equation for the
dressed vertex,
\begin{equation}
\mathcal{V}_{\alpha R}(k)
= V_{\alpha R}(k)
 + \sum_{\beta}  \int dk'\;
        {\mathcal{K}_{\alpha\beta}(k,k')
         \mathcal{V}_{\beta R}(k')
   \over   \omega_\beta(k')+E_\beta(k')-W}\,,
\label{eq4VR}
\end{equation}
and the non-resonant amplitude:
\begin{equation}
\mathcal{D}_{\alpha\gamma}(k,k_\gamma) = 
\mathcal{K}_{\alpha\gamma}(k,k_\gamma)
+  \sum_\beta \int dk'\;
        {\mathcal{K}_{\alpha\beta}(k,k')
         \mathcal{D}_{\beta\gamma}(k',k_\gamma)
   \over   \omega_\beta(k')+E_\beta(k')-W}\,,
\label{eq4D}
\end{equation}
with
\begin{equation}
    c_{\alpha R} = -{\mathcal{V}_{\alpha R}(k)\over
  \displaystyle
   W - m_R 
       + \sum_\beta\int dk\;{\mathcal{V}_{\beta R}(k){V}_{\beta R}(k)
                \over \omega_\beta(k)+E_\beta(k)-W}}\,,
\label{eq4cR}
\end{equation}
where $m_R$ is the bare mass of the resonant state.\footnote{%
Eq.~(\ref{eq4cR}) becomes more complicated if the second resonant 
state is included; see Sect.~\ref{sub:2x3q}.}

Since the kernel~(\ref{kernel}) has been rendered separable, 
equations (\ref{eq4VR}) and (\ref{eq4D}) can be solved exactly 
(i.e. to all orders) with the ansaetze
\begin{equation}
\mathcal{V}_{\alpha R}(k) = V_{\alpha R}(k) + 
  \sum_{\beta i} x^\alpha_{\beta i}\,\varphi^\alpha_{\beta i}(k)
\end{equation}
and
\begin{equation}
\mathcal{D}_{\alpha\gamma}(k,k_\gamma) =
\mathcal{K}_{\alpha\gamma}(k,k_\gamma) 
+ \sum_{\beta i} z^{\alpha\gamma}_{\beta i}\, \varphi^\alpha_{\beta i}(k)\,,
\end{equation}
where
\begin{eqnarray*}
   \varphi^{\alpha}_{\beta i}(k) &=& 
   {2m_i\over E_\beta}\,
   (\omega_\beta + \varepsilon^\beta_{i\alpha})
   {V^\alpha_{i\beta}(k)\over\omega_\alpha(k) +\varepsilon_{i\beta}^\alpha}\;
    f^i_{\alpha\beta}\,,
\\
  \varepsilon^\beta_{i\alpha} &=& 
        {m_i^2 - m_\alpha^2 - \mu_\beta^2 \over 2E_\alpha}\,.
\end{eqnarray*}
This leads to a set of linear algebraic equations for the 
coefficients 
$x$:
\begin{equation}
  \sum_{\gamma j} A^{\beta}_{\alpha i,\gamma j}\;x^\beta_{\gamma j} 
               = b^\beta_{\alpha i}\,,
\label{sys4x}
\end{equation}
and similarly for $z$ with the same $A$ matrix.
Here $i$ ($j$) refers to  $u$-channel exchange baryons 
($N$ and $\Delta$).
The expression for the $A$ matrix along with its graphical
representation as well as the RHS's $b$ are given 
in~\cite{PRC2018}.\footnote{%
In the present model the $\sigma$ terms are not included.}
Having obtained the $K$ matrix, the scattering matrix 
$T$ is obtained by solving the Heitler equation.

Increasing the interaction strength the kernel may become 
singular and the $K$ matrix acquires a pole which may be 
interpreted as a dynamically generated resonance.
In~\cite{PRC2018} we have performed the singular value 
decomposition of the  $A$ matrix~\cite{Golub96} in order to 
be able to determine the $W$-dependence of the lowest singular 
value, $w_{\mathrm{min}}$, and to study how the behavior of 
$w_{\mathrm{min}}$  is reflected in the evolution of pole(s) in 
the complex plane as the interaction strength is increased.
The main conclusion of such an analysis in our previous work 
has been that the pole corresponding to the dynamically 
generated resonance emerges well before the interaction strength 
reaches the value at which $w_{\mathrm{min}}$ touches zero.
Furthermore, the mass of the pole turns out to lie closely to 
the position where $w_{\mathrm{min}}$ reaches its minimum,
almost independently of the interaction strength.
This property persists even when $w_{\mathrm{min}}$ becomes 
negative; however, additional poles may show up at $W$ where 
$w_{\mathrm{min}}$ crosses zero.
From the corresponding eigenvectors it is possible to establish 
the main meson-baryon components of the dynamically generated 
state.

\section{\label{sec:scattering} Solving the scattering equation}

\subsection{\label{sub:no3q} No bare-baryon resonant state}

\begin{figure}[h]
\begin{center}
\includegraphics[width=85mm]{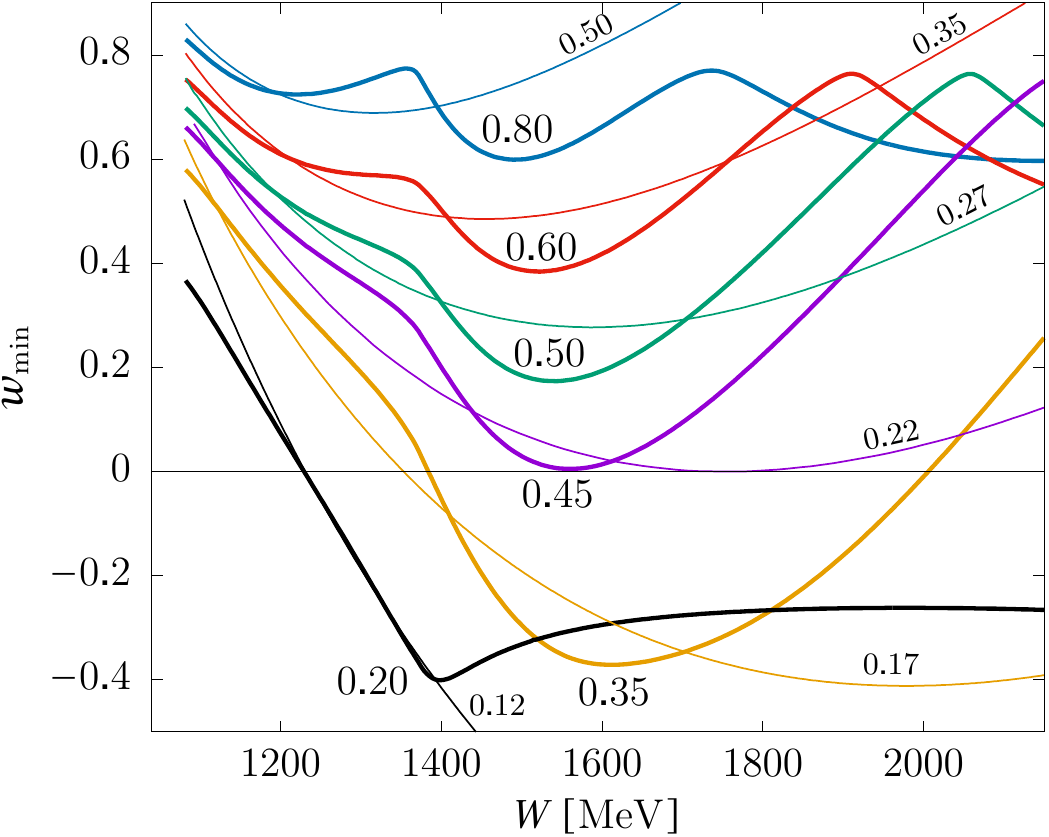}
\end{center}
\vspace{-12pt}
\caption{The behavior of $w_{\mathrm{min}}$, the lowest singular 
value of $A$ as a function of $W$ for the kernel involving $N$ 
and $\Delta$ $u$-channel exchange (thick lines) and $N$ alone 
(thin lines), for different bag radii $R$ (in fm).} 
\label{fig:wmin}
\end{figure}

We first study the case without any genuine bare baryon so
the problem reduces to solving Eq.~(\ref{eq4D}) alone.
We consider two cases; in the first one we assume only the
nucleon as the $u$-channel exchange particle and fix the $\pi NN$ 
coupling constant to the experimental value, in the second case 
we add the $\Delta$ at 1232~MeV  as the second $u$-channel exchange 
particle and fix the $\pi\Delta N$ and the  $\pi\Delta\Delta$ 
coupling constants to the quark-model values increased by 30~\%.
We vary the coupling strength by changing the bag radius.
Figure~\ref{fig:wmin} shows the behavior of $w_{\mathrm{min}}$ as a 
function of $W$ for some typical values of $R$ for the two cases.
In the first case (thinner curves) $w_{\mathrm{min}}$ touches zero for 
$R\approx0.22$~fm and crosses zero at 1232~MeV for $R=0.123$~fm.
The situation is considerably more complex in the second case
(thick curves).
For the physically interesting values around $R=0.8$~fm, 
$w_{\mathrm{min}}$ exhibits three minima at around 1200~MeV, 
1500~MeV and 2000~MeV.
For smaller $R$ the middle minimum stays close to 1500~MeV 
and touches zero for $R\approx0.45$~fm.
The zero crossing at 1232~MeV occurs for $R=0.20$~fm.

In order to obtain the scattering amplitudes we have to specify 
how to include the inelastic channel above the two pion threshold.
We assume that the decay into two pions proceeds through the 
$\pi\Delta$ intermediate state as described in~\cite{EPJ2009} and
in Appendix A of~\cite{EPJ2011}, which implies an integration over
the invariant mass of the $\pi N$ system weighted by the probability
determined in the $\pi N$ scattering in the P33 partial wave.
We assume that this probability is given by the 
Breit-Wigner mass and width of the $\Delta$.
As we shall see in the following, this assumption is consistent
for sufficiently strong coupling (small $R$) where the parameters 
of the $\Delta(1232)$ are reproduced in the same dynamical model.

\begin{figure}[h]
\begin{center}
\includegraphics[width=85mm]{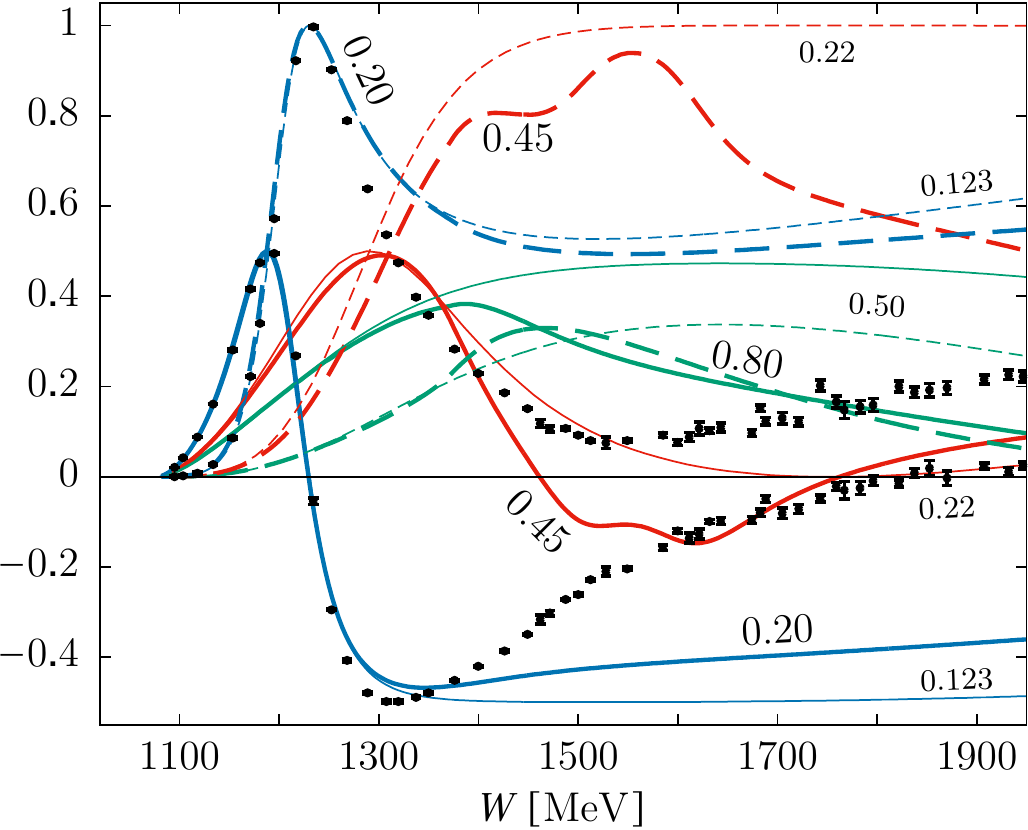}
\end{center}
\vspace{-12pt}
\caption{$T$-matrix amplitudes involving $N$ and $\Delta$ $u$-channel 
exchange (thick lines, dashes for Im$T$) 
for $R=0.8$~fm, 0.45~fm and 0.20~fm, 
and $N$ alone (thin lines,  dashes for Im$T$) 
for  $R=0.5$~fm, 0.22~fm and 0.123~fm.
Experimental data from~\cite{Arndt06,SAID}.} 
\label{fig:Tbkg}
\end{figure}

\begin{figure}[h]
\begin{center}
\includegraphics[width=85mm]{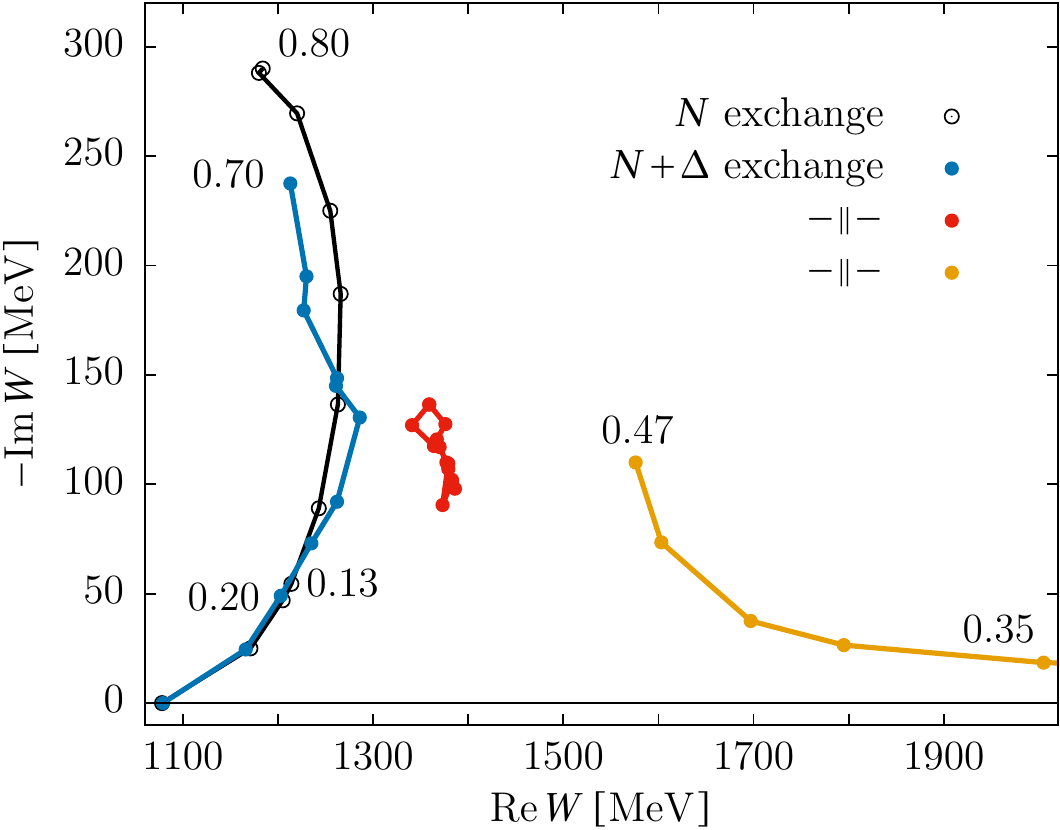}
\end{center}
\vspace{-12pt}
\caption{Evolution of poles in the complex plane as a function 
of $R$ (in fm); the single pole arising from the $u$-channel 
$N$-exchange kernel (open circles), and of the three poles from the 
$N+\Delta$ exchange kernel (full circles).} 
\label{fig:polesbkg}
\end{figure}

The resulting scattering amplitudes are displayed in 
Fig.~\ref{fig:Tbkg} for three typical bag radii.
While for larger values of $R$ the amplitudes do not show 
any visible sign of resonance, for $R=0.123$~fm 
(for the $u$-channel $N$-exchange kernel) 
and for $R=0.20$~fm ($N$ and $\Delta$-exchange) 
they perfectly fit the experimental data below 1300~MeV.
By using the L+P expansion we have been able to
follow the evolution of the pole(s) in the two cases
considered above from the (relatively) weak coupling 
towards the strong coupling regime.
Using the kernel with solely the $N$-exchange there is 
only one resonance which can be attributed to the 
dynamically generated $\Delta(1232)$.
It starts far from the real axis (i.e. with a large width), 
approaches the real axis and becomes bound for $R=0.050$~fm
(see Fig.~\ref{fig:polesbkg}).
It is interesting to notice that its pole mass remains close 
to 1200~MeV.
For $R=0.123$~fm where Re$\,T$ reaches zero at  1232~MeV, 
the pole parameters agree well with those extracted 
from experiment (see Table~\ref{tab:polesbkg}).

The situation in the case with $N$ and $\Delta$ exchange particles
is much more complex: for larger $R$ there are two poles, and
for the radius for which $w_{\mathrm{min}}$ comes close to zero,
a third pole emerges.
The lowest one exhibits a very similar evolution as the pole
in the previous case, while the second pole remains close to the 
mass 1380~MeV and width of 220~MeV, rather independently of $R$.
Since these two poles lie closely to each other, their 
determination is not very precise;
the unsmooth evolution curve in the complex plane can be 
therefore attributed to numerical instabilities.
The second pole turns out to be better determined in the
$\pi\Delta$ channel (see Table~\ref{tab:polesbkg}).
This pole can be interpreted as the progenitor of the
$\Delta(1600)$ resonance.
For smaller $R$ the mass of the third pole coincides 
with the energy at which  $w_{\mathrm{min}}$ crosses zero
the second time; it seems to have no physical interpretation
and might be an artifact of the model.

\begin{table}[h]
\caption{\label{tab:polesbkg}%
$S$-matrix pole position, modulus and phase for the $u$-channel
$N$-exchange (I) and $\Delta+N$-exchange kernel (II), 
$NN$ refers to pole determined from elastic channel and 
$\Delta\Delta$ to poles from $\pi\Delta\to\pi\Delta$.
The PDG values are taken from~\cite{PDG}.}
\begin{ruledtabular}
\begin{tabular}{ccrrr}
$R$ [fm] & Re$\,W_p$ & $-2\,{\rm Im}\,W_p$ & $|r|$ & $\vartheta$\\{}
    & [MeV]   & [MeV]   &  [MeV]    &                   \\
\colrule
PDG      & 1210 & 100 & 46    & $-46^\circ$ \\
\colrule
0.123 (I)  & 1205 &  94 & 44    & $-56^\circ$ \\
0.200 (II) & 1203 &  98 & 47    & $-55^\circ$ \\
\colrule

\colrule
PDG                   & 1510 & 270 & 25    & $-180^\circ$ \\
Ref.~\cite{L+P2014}    & 1469 & 314 & 38    & $173^\circ$ \\
\colrule
0.200 ($\Delta\Delta$) & 1376 & 255 & 10.6 & $-153^\circ$ \\
0.800 ($\Delta\Delta$) & 1379 & 219 & 15.4 & $-162^\circ$ \\
0.800 ($NN$)           & 1380 & 163 &  8.0 & $-107^\circ$ \\
\end{tabular}
\end{ruledtabular}
\end{table}

\begin{figure}[h]
\begin{center}
\includegraphics[width=85mm]{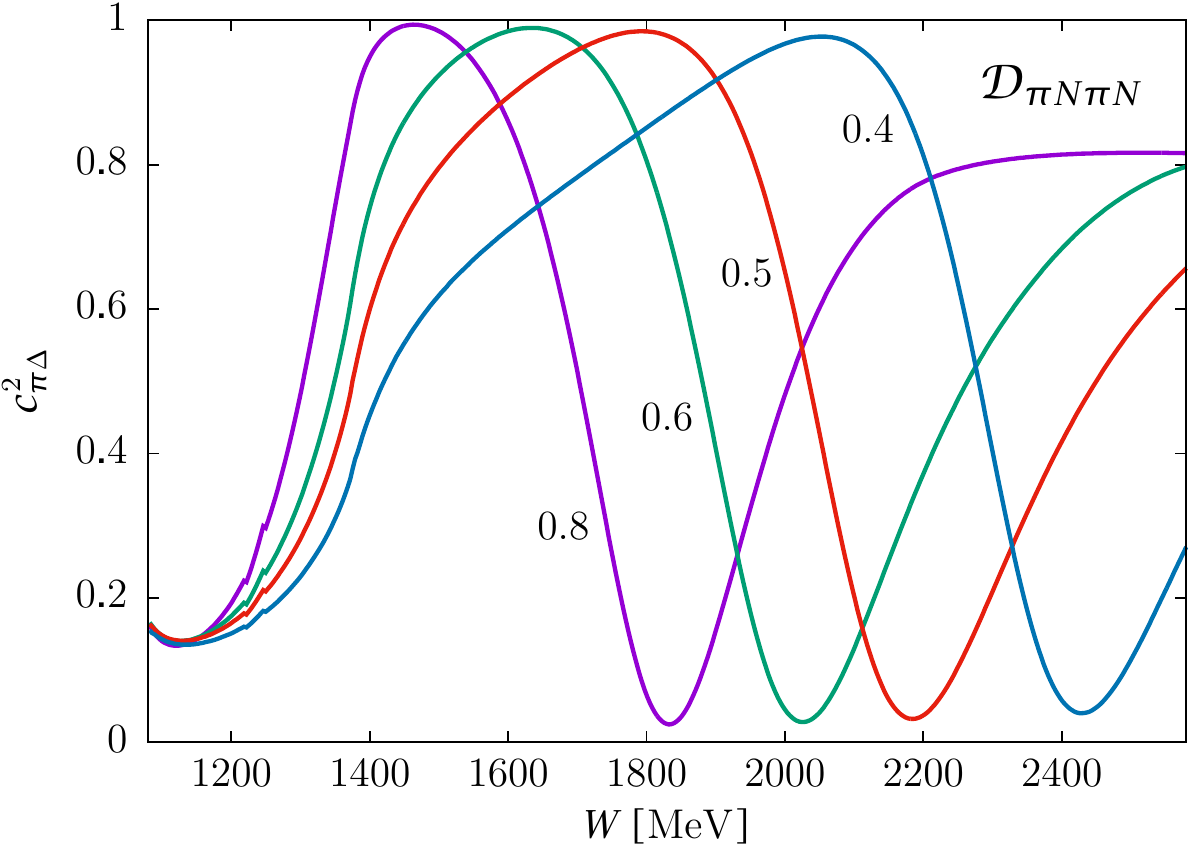}
\end{center}
\vspace{-12pt}
\caption{The probability for the $\pi\Delta$ component in
the ${\mathcal D}_{\pi N\,\pi N}$ amplitude as a function of $W$ 
for four different $R$ (in fm).} 
\label{fig:prob4piD}
\end{figure}

The singular value decomposition mentioned in the previous
section allows us to extract the probabilities for the
$\pi N$ and $\pi\Delta$ intermediate states in the scattering 
amplitudes as a function of $W$ (normalized to unity).
The probability for the $\pi\Delta$ component in elastic
scattering is displayed in Fig.~\ref{fig:prob4piD} for four
different $R$.
Similar behavior is observed also upon inclusion of the 
resonant state discussed in the next subsection.
Just about the $\pi N$ threshold the $\pi N$ component
is dominant, while in the region around and above 1400~MeV 
the  $\pi\Delta$ component starts to dominate.
This explains the constancy of width of the second resonance
which primarily depends on the width of the outgoing $\Delta$
(which is assumed to be constant in the present calculation).

\subsection{\label{sub:1x3q} Including the resonant state at 1232~MeV}

We now turn to a more realistic model introducing a genuine 
three-quark resonant state.
In order to be able to identify the most relevant degrees 
of freedom in the physically interesting region below 1700~MeV,
we work with only two channels, the $\pi N$ and the $\pi\Delta$.
We adjust the bare mass by fixing  the Breit-Wigner resonance mass 
(i.e. the zero of Re$\,T$) to 1232~MeV and further fix the bare 
$\pi N\Delta$ coupling constant to 110~\% of the quark-model value 
and the bare $\pi\Delta\Delta$ coupling constant to 55~\% in order
to (partially) compensate for the channels not taken into account.

\begin{figure}[h]
\begin{center}
\includegraphics[width=85mm]{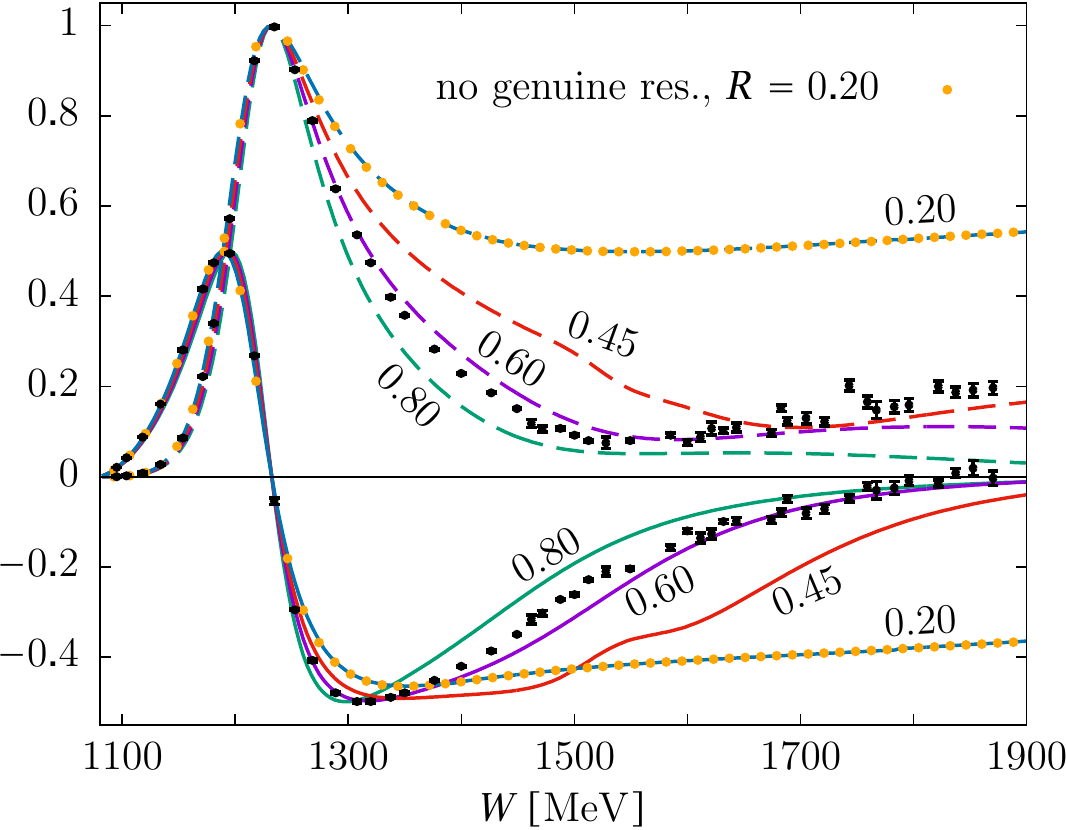}
\end{center}
\vspace{-12pt}
\caption{$T$-matrix amplitudes with and without included genuine 
resonance at 1232~MeV for $R=0.8$~fm, 0.60~fm, 0.45~fm and 0.200~fm. 
Experimental data from~\cite{Arndt06,SAID}.} 
\label{fig:TbkgND}
\end{figure}

\begin{figure}[h]
\begin{center}
\includegraphics[width=85mm]{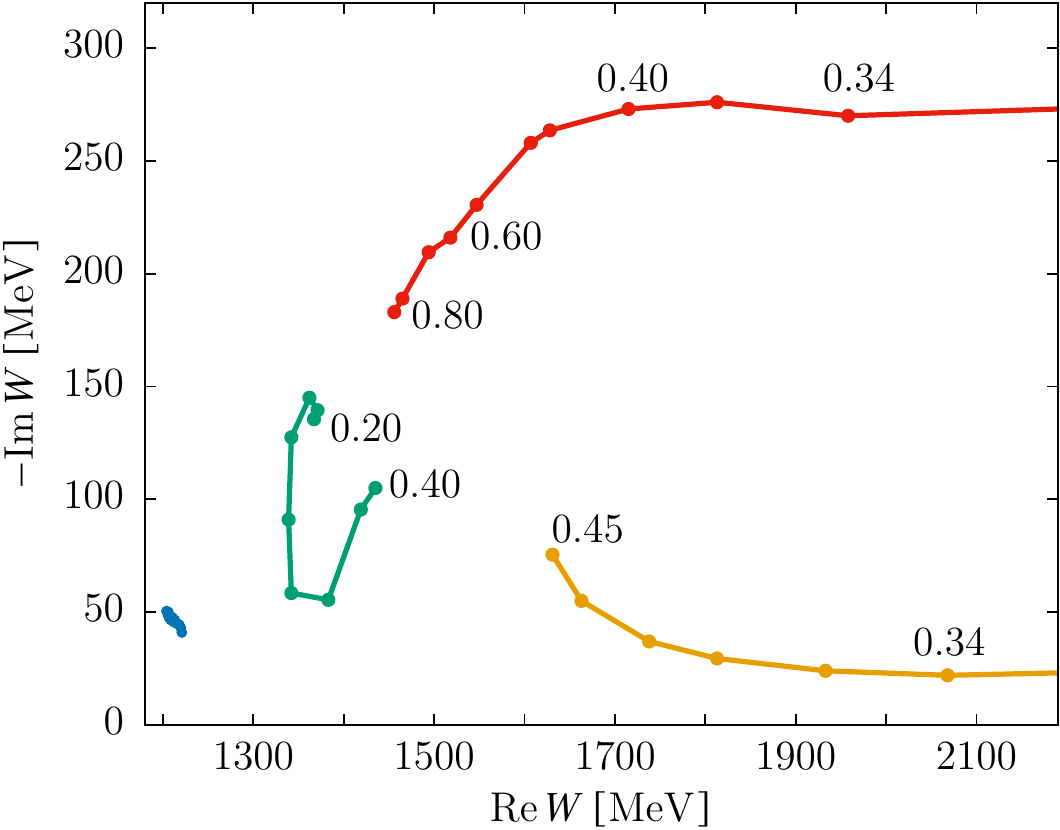}
\end{center}
\vspace{-12pt}
\caption{Evolution of poles in the complex plane as a 
function of $R$ (in fm) with included genuine resonance at 1232~MeV.} 
\label{fig:polesnd}
\end{figure}

In addition to Eq.~(\ref{eq4D}) we now solve Eq.~(\ref{eq4VR});
both equations involve the same kernel which is identical 
to the one of the purely dynamical model,
and hence the  behavior of $w_{\mathrm{min}}(W)$ coincides with
that shown in Fig.~\ref{fig:wmin}.
The scattering amplitudes are displayed in Fig.~\ref{fig:TbkgND}
for four different $R$.
The experimental amplitudes are best reproduced for $R$ between
$R=0.8$~fm and $R=0.6$~fm while at $R=0.45$~fm a structure at
around $W=1500$~ starts to become visible.
For $R=0.20$~fm the amplitudes coincide with those of the
purely dynamical model.
Analyzing the emergence and evolution of poles in the complex
plane in Fig.~\ref{fig:polesnd} we notice that the position 
of the lowest pole corresponding to $\Delta(1232)$ stays close 
to the standard PDG value and only moves toward slightly larger 
widths and smaller masses for very small $R$.
There is a second branch starting with masses around 1450~MeV and 
widths of 350~MeV evolving towards higher values for both quantities.
At smaller values, below $R$ at which  $w_{\mathrm{min}}$ crosses 
zero, a third branch emerges starting with a small residue and
evolving toward the second pole of the purely dynamical model.
There is a fourth branch essentially identical to the third
branch discussed in the previous section.   

We conclude this section by noting that the main effect of 
including a bare resonant $\Delta$ state is the mixing of
the bare state with the dynamically generated states, which 
results in pushing the dynamically generated resonance 
obtained in the previous section towards higher energies;
for larger $R$ its properties come closer to the values in 
the PDG table (see table~\ref{tab:poles1600}) and it can be 
identified as the $\Delta(1600)$ resonance.

\begin{table}[h]
\caption{\label{tab:poles1600}%
  $S$-matrix pole position, modulus and phase
  with one resonant state for $R=0.8$~fm, 0.6~fm, 0.2~fm,
  and with an additional resonant state with a bare mass 2.2~GeV 
  and 2.0~GeV.
  The PDG values are taken from~\cite{PDG}.}
\begin{ruledtabular}
\begin{tabular}{ccrrr}
$R$ [fm] & Re$\,W_p$ & $-2\,{\rm Im}\,W_p$ & $|r|$ & $\vartheta$\\{}
    & [MeV]   & [MeV]   &  [MeV]    &                   \\
\colrule
PDG                & 1510 & 270 & 25    & $-180^\circ$ \\
Ref.~\cite{L+P2014} & 1469 & 314 & 38    & $173^\circ$ \\
\colrule
0.800              & 1449 & 350 & 26.3 & $174^\circ$ \\
0.600              & 1508 & 427 & 50.0 & $-165^\circ$ \\
0.200              & 1367 & 271 & 12.1 & $-160^\circ$ \\
\colrule
$m_{\Delta^*}=2.2$~GeV & & & & \\
\colrule
0.800              & 1453 & 360 & 44   & $-174^\circ$ \\
0.600              & 1570 & 397 & 66   & $-166^\circ$ \\
\colrule
$m_{\Delta^*}=2.0$~GeV & & & & \\
\colrule
0.800              & 1452 & 347 & 46   & $-179^\circ$ \\
0.600              & 1631 & 340 & 68   & $-140^\circ$ \\
\end{tabular}
\end{ruledtabular}
\end{table}

\subsection{\label{sub:2x3q} Including the second resonant state}

We finally consider the inclusion of the second genuine three-quark 
resonant state in which one quark is promoted to the $2s$ state.
In our previous work on the $N(1440)$ we have found a strong 
mixing of such a configuration with the dynamically generated 
state; it is therefore important to check whether such a mixing 
plays a sizable role also in the P33 partial wave.

As discussed in the Introduction we expect that a bare state 
with such a configuration does not exist below 2000~MeV.
We therefore consider two possible bare masses of
2200~MeV and 2000~MeV.
We used the pion coupling constants obtained in the
Cloudy Bag Model without any further adjustment:
for $\pi NR$ and $\pi\Delta R$ the constants are 46~\% of 
the corresponding quark-model values for the $(1s)^3$ 
configuration. 
We assume the same bag radius for both states.

The calculation proceeds 
by introducing an additional term
$c_{\gamma R^*}{\cal V}_{\alpha R^*}(k)$ in (\ref{eq4chi}) with
${\cal V}_{\alpha R^*}(k)$ satisfying an analogous equation 
to Eq.~(\ref{eq4VR}).
The equations for $c_{\gamma R}$ and $c_{\gamma R^*}$, however, 
become more complicated due to mixing of the bare resonances
through pion loops, as described in~\cite{PRC2018}.
We fix again the Breit-Wigner mass of the lower resonance 
to 1232~MeV by adjusting the bare $\Delta$ mass; this procedure
works, provided that the bare $\Delta$ mass does not come too
close to the bare mass of the upper states, which occurs
around $R=0.45$~fm.
The resulting Breit-Wigner mass of the upper resonance
remains close to its bare value even for smaller $R$ where
the two bare state strongly mix.

Figure~{\ref{fig:TNDR} displays the $T$-matrix amplitude with 
and without including the second resonant state for the bag 
radii for which the amplitude agrees best with experiment.
While the imaginary part of the amplitude is improved 
in the intermediate regime by introducing the second state,
a typical resonant behavior emerging around 2000~MeV for 
the bare mass of 2000~MeV is not supported by experiment.
It seems therefore  that such a low mass is ruled out.
Adding further channels may enhance the amplitude
and improve the agreement 
and may eventually support a bare mass as low as 2200~MeV.
We shall nonetheless include both bare states in our
further analysis.

\begin{figure}[h]
\begin{center}
\includegraphics[width=85mm]{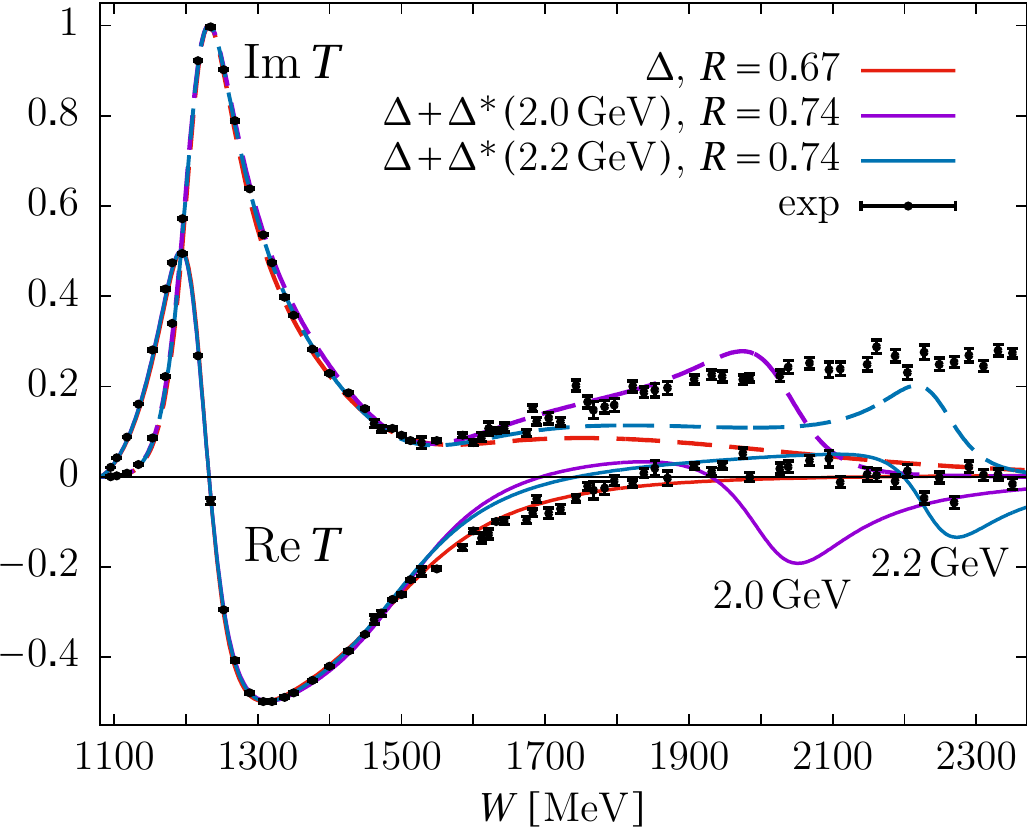}
\end{center}
\vspace{-12pt}
\caption{$T$-matrix amplitude with the included resonant $\Delta$
state  for $R=0.67$~fm, and the amplitude with the included second 
state at bare masses 2000~MeV and 2200~MeV for $R=0.74$~fm.
Experimental data from~\cite{Arndt06,SAID}.} 
\label{fig:TNDR}
\end{figure}

The evolution of the poles pertinent to the second and the third 
resonance is compared to the evolution of the second resonance 
discussed in the previous subsection in Fig.~\ref{fig:polesNDR}.
We notice that the presence of the new resonant state 
affects little the properties for larger $R$ but already
for $R=0.60$~fm the mass of the resonance is further 
increased with respect to the purely dynamical resonance
as well as the resonance with a single resonant state
(see also Table~\ref{tab:poles1600}).
The third resonance pole, starting on the real axis at the bare 
mass of the resonant state, evolves towards large $-{\rm Im}\,W$ 
almost in a straight line and does not bend toward lower 
masses as in the case of the $N(1440$) resonance.
This indicates that the excited quark configuration 
plays a rather insignificant role in the formation of the
$\Delta(1600)$ resonance (in the physically sensible range of radii).

\begin{figure}[h]
\begin{center}
\includegraphics[width=85mm]{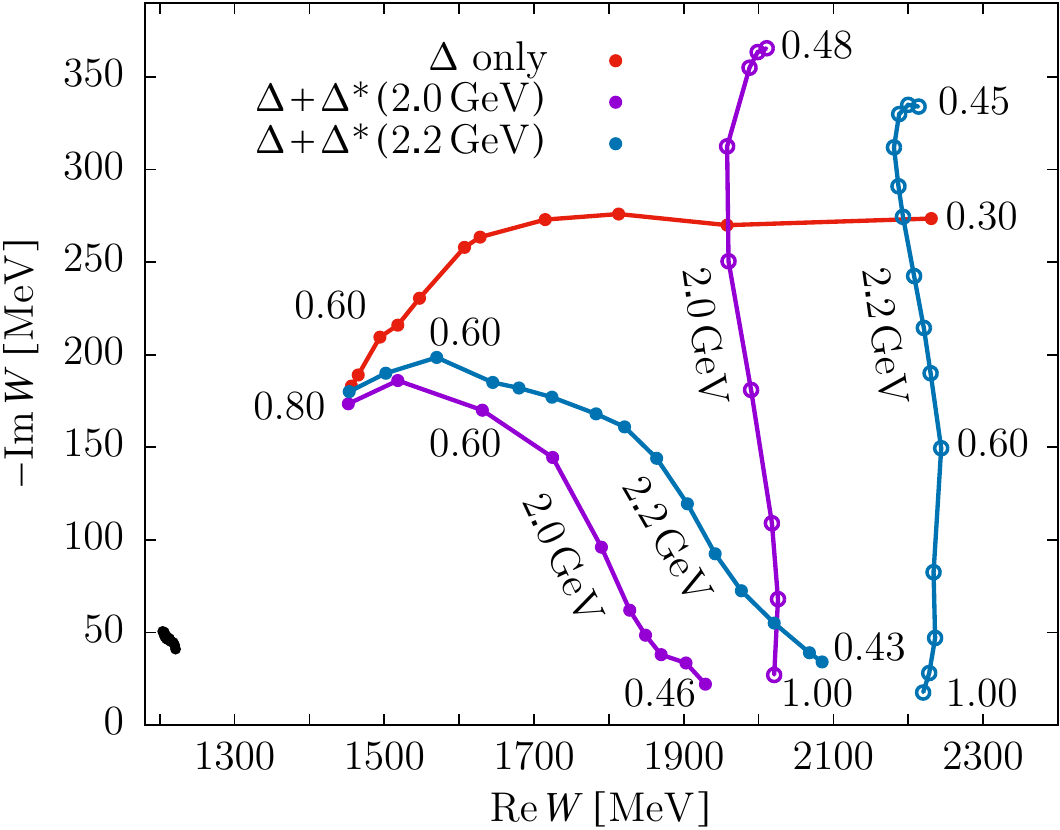}
\end{center}
\vspace{-12pt}
\caption{Evolution of the poles pertinent to the second
resonance (full circles) and the third resonance (open circles)
in the complex plane, as a function of $R$ (in fm).} 
\label{fig:polesNDR}
\end{figure}

\section{\label{sec:electro} Electroproduction amplitudes}

Electroproduction is an important tool to study the
resonance structure. 
In particular, the extracted $Q^2$-dependence of helicity 
amplitudes may reveal the spatial distribution of quark 
and meson degrees of freedom.
As an example, let us mention our early calculation
of the helicity amplitudes in the case of the $\Delta(1232)$
resonance~\cite{PLB1996} in which we have been able to
disentangle the quark and the pion degrees of freedom
and shown that the latter strongly dominate the
$E_{1+}$ production amplitude; similarly, studying the
structure of the $N(1440)$ resonance~\cite{EPJ2009} 
we have been able to explain the zero crossing of the
$A_{1/2}$ amplitude by its transition from the pion-dominated
region at low $Q^2$ to the region dominated by quarks
at larger $Q^2$.
This type of calculation is based on the assumption that 
it is  possible to extract the resonant state from (\ref{PsiH}) 
and calculate the electro-magnetic (EM) part of the 
electroproduction amplitude by evaluating the corresponding 
EM multipole between the ground and the resonant state.
Such an approach is justified if the resonance is
sufficiently narrow and can be separated from the background
and possible neighboring resonances.
This is certainly not fulfilled in the case of $\Delta(1600)$.
It seems that the only physically sensible way to obtain the
helicity amplitudes is to extract them from the
electroproduction amplitudes at the $S$-matrix pole for 
different values of $Q^2$.

The calculation is organized as follows:
the $K$ matrix acquires a new channel, $\gamma N$,
and the corresponding matrix elements  are calculated
from (\ref{eq4K}) with the EM interaction,
\begin{equation}
V^\gamma_\mu(\vec{k}_\gamma)
  = {\mathrm{e}_0\over\sqrt{2\omega_\gamma}}
    \int d\vec{r}\,\vec{\varepsilon}_\mu\cdot\vec{j}(\vec{r})
    \,e^{i\svec{k}_\gamma\cdot\svec{r}}\,,
\label{Vgamma}
\end{equation}
replacing the strong one.
Here  $\vec{k}_\gamma$ and $\mu$ are the momentum 
and the polarization of the incident photon, and
the current involves the quark and the pion part:
$$
  \vec{j}(\vec{r})  =
  \bar{\psi}\vec{\gamma}({\textstyle{1\over6}} + \half\tau_0)\psi
  + i \sum_t t \pi_t(\vec{r})\vec{\nabla}\pi_{-t}(\vec{r})\;.
$$
Only $M1$ and $E2$ multipoles contribute in the $P33$ partial wave.
The $M_{1+}$ and $E_{1+}$ amplitudes are calculated and the 
respective residues are determined by using the L+P expansion.
From these residues the helicity amplitudes 
(photodecay amplitudes) $A_h^{\mathrm{pole}}$ are calculated 
as~\cite{Tiator13}:
\begin{equation}
 A_h^{\mathrm{pole}} = \sqrt{16\pi\,k_\pi\,M_p\over3 k_\gamma m_N 
                    {\rm Res}_{\pi N}}\;{\rm Res}\mathcal{A}^h_{1+}\,,
\label{M2A}
\end{equation}
valid for the P33 partial wave.
Here $k_\pi$ and $k_\gamma$ are the pion and photon momenta
evaluated at the pole, $M_p$ is the mass of the resonance and
 ${\rm Res}_{\pi N}$ the elastic $\pi N$ residue;
$\mathcal{A}^{1/2}_{1+}=-\half\left(M_{1+}+3E_{1+}\right)$ and 
$\mathcal{A}^{3/2}_{1+}=-{\sqrt3\over2}\left(M_{1+}-E_{1+}\right)$.

In the first step we calculate the photoproduction amplitudes
and compare them to the residues obtained by \v{S}varc et 
al.~\cite{L+P2014a} for $\gamma N\to \pi N$ and of Sokhoyan 
et al.~\cite{Sokhoyan15} for $\gamma N\to \pi\Delta$; 
see Table~\ref{tab:photo}.

\begin{table}[h]
\caption{\label{tab:photo}%
Photoproduction residues at the pole of $\Delta(1600)$.
The values of the modulus should be divided by 1000.}
\begin{ruledtabular}
\begin{tabular}{ccccc}
           & $|{\rm Res}E_{1+}|$ & $\theta_E$  
           & $|{\rm Res}M_{1+}|$ & $\theta_M$ \\
\colrule
$\gamma N \to \pi N$  &&&& \\
\colrule
Ref.~\cite{L+P2014a}  &    $0.44$ & $127^\circ$     
                      &    $2.53$ & $-149^\circ$ \\
$R=0.8$~fm            &    $0.30$ & $143^\circ$     
                      &    $2.02$ & $173^\circ$ \\
$R=0.6$~fm            &    $0.49$ & $115^\circ$     
                      &    $4.46$ & $122^\circ$ \\
\colrule
$\gamma N \to \pi\Delta$  &&&& \\
\colrule
Ref.~\cite{Sokhoyan15}   & $2\pm1$      & $30^\circ\pm30^\circ$ 
                         & $12\pm3$     & $ 65^\circ\pm25^\circ$ \\
$R=0.8$~fm               & $0.41$       & $158^\circ$     
                         & $6.51$       & $119^\circ$ \\
$R=0.6$~fm               & $0.67$       & $158^\circ$     
                         & $13.3$       & $127^\circ$ \\
\end{tabular}
\end{ruledtabular}
\end{table}

\begin{figure}[h]
\begin{center}
\includegraphics[width=85mm]{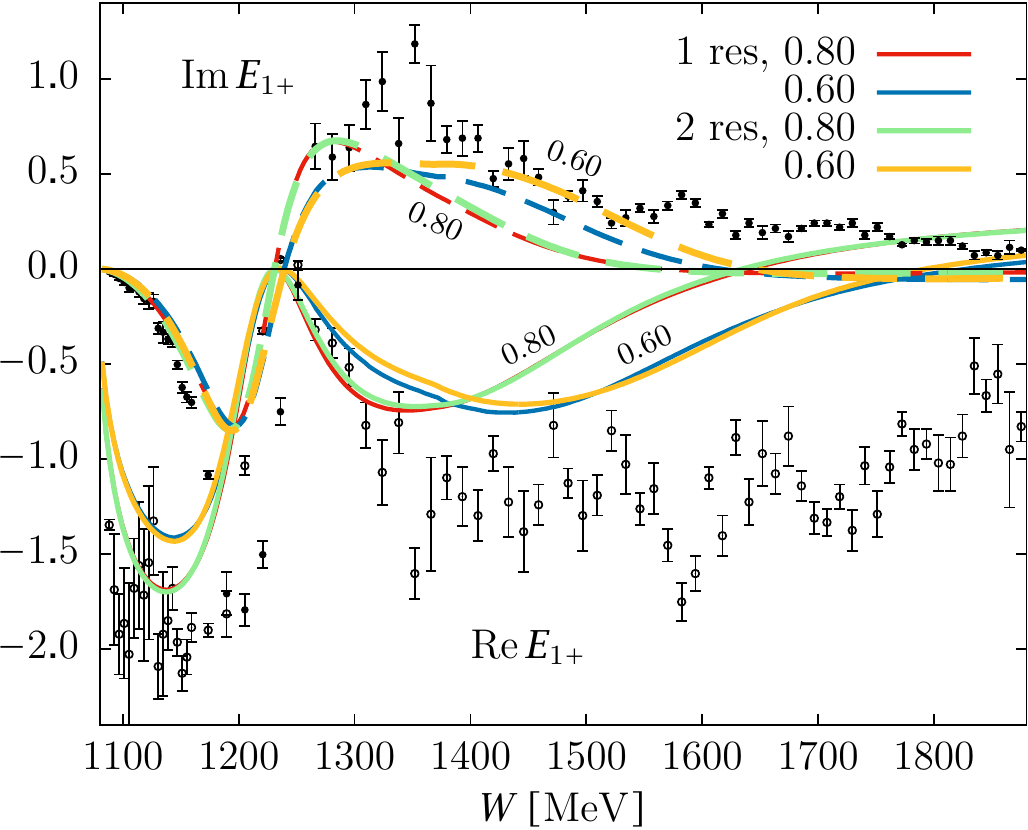}
\end{center}
\vspace{-12pt}
\caption{Photoproduction $E_{1+}$ amplitude (in mfm) for $R=0.6$~fm 
  and 0.8~fm evaluated in the model with one or two resonant 
  three-quark states, the second one at $m_{\Delta*}=2.2$~GeV.
  Experimental data from~\cite{Arndt06,SAID}.} 
\label{fig:E1p}
\end{figure}

In the next step we calculate the $Q^2$-dependence of the helicity 
amplitudes $A_{1/2}$ and $A_{3/2}$ at the pole
by first calculating the electroproduction amplitudes as
a function of $W$ at finite $Q^2$ and then determine the residue of 
the pole for each $Q^2$ separately by using the L+P expansion.%
\begin{figure*}[t]
\begin{center}
\includegraphics[width=170mm]{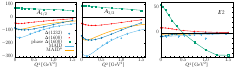}
\end{center}
\vspace{-12pt}
\caption{Helicity amplitudes 
  (units $10^{-3}{\rm GeV}^{-1/2}$, phase in degrees)
  at the $S$-matrix pole of the $\Delta(1232)$ and $\Delta(1600)$ 
  resonances for $R=0.8$~fm, and the corresponding multipole 
  amplitude  $E2$,
  compared to the phenomenological parameterization~\cite{MAID2007}.}
\label{fig:Ahs}
\end{figure*}
The resulting amplitudes are displayed in Fig.~\ref{fig:Ahs}
and compared to the phenomenological amplitudes of the
MAID2007 analysis~\cite{MAID2007} evaluated at $W=1232$~MeV
and at 1470~MeV.
One should keep in mind that the latter amplitudes have been 
evaluated by assuming a Breit-Wigner behavior for the resonance, 
while ours are extracted from the pole residue.
It is known that the quark contribution to the magnetic amplitudes 
is considerably underestimated in our model, particularly at
smaller $Q^2$, due to the fact that the spinors are limited
to the interior of the bag. 
We expect that this effect is present also in the amplitudes
pertinent to the $\Delta(1600)$ resonance.
The helicity amplitudes roughly  follow the same trend
for both resonances, similar to the MAID amplitudes.
At low $Q^2$ there is, however, a substantial difference
due to the $E2$ multipole, which is considerably
larger than in the phenomenological parameterization; 
as a result, the $A_{1/2}$ amplitude at the photon point becomes 
almost equal to the $A_{3/2}$.
Taking into account the simplicity of the quark model embedded 
into our coupled-channels framework as well as large uncertainty 
of the experimental data regarding this resonance, our values 
compare favorably to the values from Refs.~\cite{Sokhoyan15,BoGa12}
(see Table~\ref{tab:helicity}).
Since the $E2$ multipole contribution originates entirely
from the pion current, this effect is a strong signal for the
important role of the pion cloud in the $\Delta(1600)$ and
supports our picture of the resonance.

Let us stress that our model reproduces reasonably well 
the $E_{1+}$ photoproduction amplitude (Fig.~\ref{fig:E1p}) 
in the physically relevant range of $W$, which speaks in 
favor of our model.
The figure also shows that the presence of the second
baryon state at 2.2~GeV has little effect, particularly
for larger $R$.

\begin{table}[h]
\caption{\label{tab:helicity}%
Helicity amplitudes at the photon point in units of 
$10^{-3}{\rm GeV}^{-1/2}$.}
\begin{ruledtabular}
\begin{tabular}{ccccc}
           & $|A_{1/2}^{\mathrm{pole}}|$ & $\theta_{1/2}$  
           & $|A_{3/2}^{\mathrm{pole}}|$ & $\theta_{3/2}$ \\
\colrule
Ref.~\cite{Sokhoyan15}  & $53\pm10$    & $(130\pm15)^\circ$
                        & $55\pm10$    & $(152\pm15)^\circ$ \\
Ref.~\cite{Roenchen14}  & $193$        & $15^\circ$
                        & $254$        & $175^\circ$       \\
Ref.~\cite{BoGa12}      & $53\pm10$    & $(130\pm25)^\circ$
                        & $41\pm11$    & $(165\pm17)^\circ$ \\
Ref.~\cite{Kamano13}    & $72$         & $-109^\circ$
                        & $136$        & $-98^\circ$ \\
$R=0.8$~fm              & $44$         & $-104^\circ$     
                        & $47$         & $-89^\circ$ \\
$R=0.6$~fm              & $68$         & $22^\circ$     
                        & $79$         & $25^\circ$ \\
\end{tabular}
\end{ruledtabular}
\end{table}

\section{Conclusion}

We have studied the mechanism of resonance formation 
in the P33 partial wave in a model including only the
$\pi N$ and the $\pi\Delta$ channels.
The L+P expansion has been used to extract the 
$S$-matrix resonance-pole parameters.
We have shown that assuming only the $u$-channel exchange 
diagrams, the system supports two resonances of dynamical 
origin, the lowest one  with a pole mass around 1200 MeV, 
dominated  by the $\pi N$ loops, and the second one, 
dominated by the $\pi\Delta$ loops, with a mass slightly 
below 1400~MeV.
For a sufficiently large cutoff parameter, the position
and the residue almost perfectly agree with the PDG values,
yet the corresponding size of the quark core appears 
to be much too small to make such a model realistic.

Including a genuine three quark resonant state in the 
$s$-channel it becomes possible to reproduce sufficiently 
well the scattering data in the intermediate energy region 
by using physically sensible values for the  cutoff. 
The properties of the $\Delta(1232)$ are well reproduced and, 
furthermore, the second dynamically generated resonance is 
pushed toward a somewhat higher pole mass of around 1500~MeV 
and acquiring the width which agrees with still rather 
uncertain PDG values for the $\Delta(1600)$ resonance.

We have checked whether the inclusion of the $s$-wave $\sigma$ 
meson as well as the quark configuration with one quark excited 
to the $2s$ state --- which has turned out to play a dominant 
role in the case of the $N(1440)$ resonance --- 
may change the above picture.
The $\sigma\Delta$ channel, as the counterpart of the $\sigma N$
channel in the P11 partial wave, starts to influence the results
only above 1700~MeV, while the evolution of the pole stemming
from the leading $(1s)^22s$ configuration is well separated from 
the pole evolution pertinent to the second dynamically generated 
resonance.
The excited quark core configuration could eventually be the
dominant ingredient in one of the higher P33 resonances.

Electroproduction in the energy region below $\approx 1700$~MeV, 
in particularly the extraction of helicity amplitudes at 
finite $Q^2$, seems to be the most decisive test to confirm 
the validity of our picture of the $\Delta(1600)$ resonance.
Our model predicts a relatively strong contribution from 
the $E2$ multipole originating solely from the photon 
interaction with the pion cloud and dominating at small 
$Q^2$ due to the large extent of the pion field.
As a result, $A_{1/2}$ is enhanced, while $A_{3/2}$ is
diminished with respect to their ratio of $\sqrt3$ assuming
$M1$ dominance, and the two amplitudes in fact become 
comparable in size.
Furthermore, we predict that they drop smoothly to zero at 
large $Q^2$ and do not exhibit a zero crossing as is the case 
with the $A_{1/2}$ amplitude in the Roper resonance where 
the quark $(1s)^22s$ configuration is strong and produces
a contribution with the opposite sign with respect to the
pion cloud contribution.

We conclude that the $\Delta(1600)$ is perhaps the most clean
example of a dynamically generated non-strange resonance in
the second and third resonance regions.

\end{document}